\documentclass[prl,aps,twocolumn,showpacs]{revtex4}

\usepackage{amsfonts}
\usepackage{epsfig}
\usepackage{bm}

\begin{document}

\title{A non-abelian spin-liquid in a spin-1 quantum magnet}

\author{Tarun Grover} 
\affiliation{Department of Physics, UC Berkeley, Berkeley, California 94720}

\author{T. Senthil}
\affiliation{Department of Physics, Massachusetts Institute of Technology, Cambridge, Massachusetts 02139}

\begin{abstract}
We study a time-reversal invariant non-abelian spin-liquid state in an $SU(2)$ symmetric spin $S=1$ quantum magnet on a triangular lattice. The spin-liquid is obtained by quantum disordering a non-collinear nematic state. We show that such a spin-liquid cannot be obtained by the standard projective construction for spin-liquids. We also study phase transition between the spin-liquid and the non-collinear nematic state and show that it cannot be described within Landau-Ginzburg-Wilson paradigm.
\end{abstract}


\maketitle

The concepts of `order parameter'  and `symmetry breaking' have been extremely successful for classifying various phases of matter and their phase transitions at finite temperature. Interestingly, last twenty years have provided us many examples where these notions break down in the context of zero temperature quantum phases and phase transitions and where one needs a totally different framework. In particular, the concepts such as `fractionalization' and `topological order' have been shown to be very useful to describe rich physics of various quantum hall systems, frustrated magnets and many other strongly correlated systems \cite{wen_book}.

Our current description of topological order draws heavily on the projective construction using various slave boson/fermion techniques \cite{wen_book}. In this paper we present an example of a non-abelian quantum spin liquid with bosonic spinons where the Schwinger boson construction fails to capture even the mean field state. The spin-liquid is obtained by quantum disordering a non-collinear nematic state in a spin $S=1$ system on a triangular lattice. Interestingly, the low energy excitations in the spin-liquid have non-abelian statistics! To our knowledge, this is one of the few known examples of a time-reversal invariant phase of matter that supports non-abelian excitations \cite{kitaev, levinwen05}. Further, the phase transition between the nematic phase and the spin-liquid is also exotic and has a very large value of the critical exponent $\eta$ associated with correlations of nematic order parameter.

A general Hamiltonian describing a spin $S= 1$ quantum magnet on an isotropic triangular lattice takes the form
 
\begin{equation}
 H = J \sum_{<ij>} \, S_i.S_j + K\, \sum_{<ij>} \left( S_i.S_j\right)^2  \label{ham}
\end{equation}

Additionally, $H$ may have other short-ranged interactions consistent with $SU(2)$ spin symmetry such as multiple ring-exchange or second nearest neighbor Heisenberg exchange. 

Consider the Hamiltonian $H$ for $K > J > 0$. The ground state of this Hamiltonian in this parameter regime has a three-sublattice nematic order where the nematic directors on the three sublattices $A$, $B$, and $C$ of the triangular lattice are orthogonal to each other (say along $\hat{x}, \hat{y}$ and $\hat{z}$ respectively) \cite{fath95}. As argued by Tsunetsugu and Arikawa \cite{tsunet05}, such a state may explain many of the features\cite{nigas} of the triangular lattice magnet $Ni Ga_2 S_4$, in particular, the lack of any dipole moment $\left\langle \bm{S} \right\rangle$, low temperature specific heat $C(T) \sim T^2$ and finite spin susceptibility at $T = 0$. The directors of the nematic correspond to `hard-axes' i.e. the spins on the three sublattices fluctuate in the plane perpendicular to their respective directors such that average value $\left\langle \bm{S(r)} \right\rangle = 0$.  Such a state breaks spin-rotation symmetry while preserving the time-reversal invariance. In terms of the the spin operators, the nematic order parameter at site $\bm{r}$ could be described as a rank-two tensor $Q_{\mu \nu}(\bm{r}) = \frac{1}{2}\left\langle S_{\mu}(\bm{r}) S_{\nu}(\bm{r}) + S_{\nu}(\bm{r}) S_{\mu}(\bm{r}) \right\rangle  - \frac{2}{3} \delta_{\mu \nu} $. The director at a site $\bm{r}$ is along the eigenvector of $Q_{\mu \nu}$ that corresponds to the zero eigenvalue.

The above ground state doesn't preserve any continuous subgroup of the original $SO(3)$ symmetry of the Hamiltonian in eqn. \ref{ham}. Thus the low-energy fluctuations around the ground state consist of three goldstone modes. Interestingly, the ground state is invariant under the discrete subgroup $D_2 \equiv R^x_{\pi}, R^y_{\pi}, R^z_{\pi}$ of $SO(3)$ where $R^a_{\pi}$ ($a = x,y,z$) corresponds to a global $\pi$ rotation of all spins about the three orthogonal axes $x, y, z$. Therefore the order-parameter manifold $\mathcal{M} = SO(3)/D_2$. This identification of the order-parameter manifold allows one to characterize the non-trivial topological excitations out of the ground state. We recall that in two spatial dimensions the fundamental group $\pi_1(\mathcal{M})$ of the order parameter manifold $\mathcal{M}$ is directly related to the combination law for the physical point defects \cite{mermin79}. More precisely, the defects are classified by the conjugacy classes of the fundamental group. To calculate the fundamental group of $\mathcal{M} = SO(3)/D_2$, one notes that the lift of $D_2$ in $SU(2)$ is the eight element non-abelian Quaternion group $Q$. Thus $\mathcal{M}$ is homeomorphic to $SU(2)/Q$. Since $SU(2)$ is simply connected while $Q$ is discrete, using the fundamental theorem on the fundamental group, $\pi_1(\mathcal{M}) = Q$ \cite{mermin79}. As we will shortly see that the two-dimensional representation of $Q$ in terms of Pauli matrices would be most relevant for our purposes. In this representation the five conjugacy classes of $Q$ are given by:

\begin{eqnarray}
 C_0 & = & \left\lbrace \mathbb{I} \right\rbrace , \,\, \bar{C_0} = \left\lbrace \mathbb{-I} \right\rbrace \nonumber \\
C_x & = & {\pm i \sigma_x}, \,\, C_y = {\pm i \sigma_y}, \,\, C_z = {\pm i \sigma_z} \label{quater_conj}
\end{eqnarray}

The class $C_0$ corresponds to the trivial class i.e. no defect. The class $\bar{C_0}$ corresponds to a $360^{\circ}$ disclination in two of the nematic directors while the class $C_a$ ($a = x, y, z$) corresponds to defects where there is a $180^\circ$ disclination in all but $a$-axis directors. 

\begin{figure}[tb]
\centerline{
 \includegraphics[scale = 0.7]{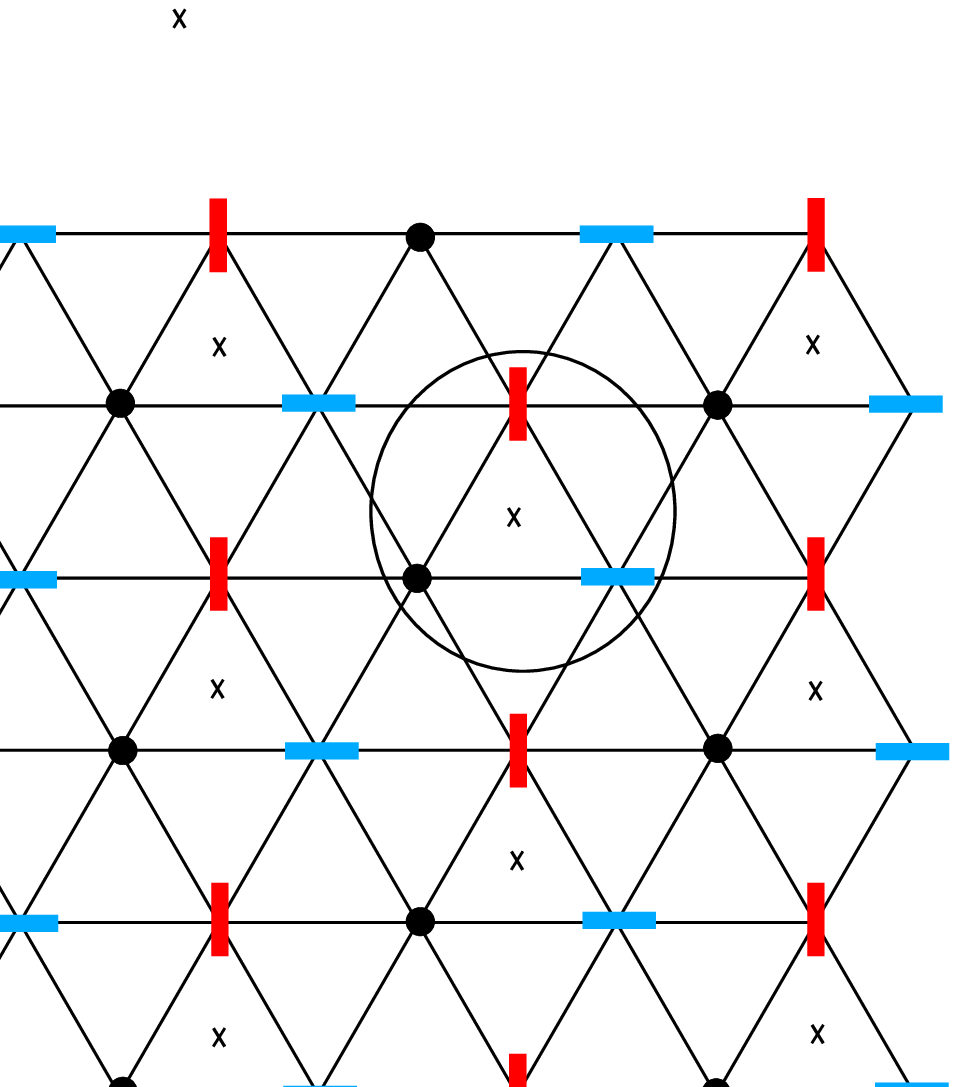}
}
\caption{The ordered state of the three sublattice nematic state. The directors on the three sublattices are shown as three orthogonal differently colored rods (black, red and blue). As discussed in the text, one may also define the order parameter space in terms of spinons $z$ which reside at the vertices of a bigger auxiliary triangular lattice. We take the vertices of this auxiliary triangular lattice (marked as $`\times'$) to lie at the centroid of every third triangular plaquette of the original lattice.}
\label{nem_tri}
\end{figure}

We are interested in constructing a $T = 0 $ spin-liquid state obtained by quantum disordering the nematic state.  Since proliferating topological defects in quantum magnets often leads to breaking of various symmetries in the paramagnet state, one of the simplest ways to obtain a spin-liquid is to destroy the nematic state \textit{without} proliferating any topological defects. To implement this, we use an effective lattice model formulated as a gauge theory analogous to the formulation of classical spin-nematics as a lattice field theory of a $Z_2$ gauge field coupled to a vector field \cite{lammert93}.
 
The order parameter space $\mathcal{M} = SO(3)/D_2$ is equivalent to a set of orthogonal axes $\bm{n_1, n_2}$ and $\bm{n_3}$ at each vertex of a bigger triangular lattice with the identification $\bm{n_a} \equiv R \, \bm{n_a}$ where $R$ is an element of the $D_2$ group. The vertices of this new triangular lattice could be taken as the centroids of the triangular plaquettes of the original triangular lattice (figure 1). Equivalentally, using the identification $SO(3)/D_2 \equiv SU(2)/Q$, it could be described as a quaternion gauge-matter theory whose imaginary time action is:

\begin{eqnarray}
 S & = & -t \sum z^{*}_i \, q_{ij} z_j + \kappa \left[  \sum_{\triangle} Tr \, \left( q_{ij}\, q_{jk}\, q_{ki}\right) \right. \nonumber \\
 & &  + \left. \sum_{\Box} Tr \, \left( q_{ij}\, q_{jk}\, q_{kl}\, q_{li}\right) \right]  + S_B \label{action_z} 
\end{eqnarray}

Here $z = \left[ z_{\uparrow}, z_{\downarrow} \right]^T $ is a two component spinor with the constraint $z^{\dagger} z = 1$ which is minimally coupled to a quaternion gauge field $q_{ij}$. $z$'s and $q$'s live at the vertices and the links respectively of a stacked triangular lattice. The first term in the action $S$ is the kinetic energy term for the spinons $z$ while the second term is the kinetic energy term for the gauge fields at a spatial ($\triangle$) or space-time ($\Box$) plaquette. The term $S_B$ is the Berry phase term associated with topological defects. Since the topological defects are gapped in all the phases considered in this paper, $S_B$ can effectively be ignored for our purposes. The original $SO(3)$ symmetry of the spin-Hamiltonian $H$ is realized as a $SU(2)$ symmetry acting on the spinor $z$. This also indicates how to relate the spinor $z$'s to the axes $\bm{n_1, n_2}$ and $\bm{n_3}$. Let us define an $SU(2)$ matrix $U$ build from $z$'s:

\begin{equation}
 U = \left( \begin{array}{cc} z_{\uparrow} & z^{*}_{\downarrow} \\ z_{\downarrow} & -z^{*}_{\uparrow}    \end{array} \right) \label{defn_U}
\end{equation}

The $3 \times 3$ matrix $R$ build from $U$

\begin{equation}
 R^{ab} = \frac{1}{2} \, tr\left( U^{\dagger} \sigma^{a} U \sigma^{b}\right) 
\end{equation}

has $\bm{n_1, n_2}$ and $\bm{n_3}$ as its first, second and third columns respectively. Thus, all three directors $\bm{n_a}$'s are quadratic in $z$'s and hence the nematic order parameter $Q_{\mu \nu}$ is of fourth degree in $z$'s.

Let us describe the phase diagram corresponding to the action $S$ in eqn. \ref{action_z}. First consider $S$ at finite temperature. Since a continuous symmetry cannot be broken in two dimensions with short-range interactions, no symmetry broken phases can exist at any non-zero temperature. Further, since a pure quaternion gauge theory in two spatial dimensions is confining at any non-zero temperature, in terms of the original spin model only a paramagnet phase without any topological order can exist at $T > 0$.

Next consider the phase diagram at zero temperature. Clearly for $t \gg \kappa$, the spinons would condense yielding a three sublattice nematic state. As discussed above, this phase would have three goldstone modes. The non-abelian fluxes through plaquettes would exist as excitations in this  phase and correspond to the non-abelian disclinations in eqn. \ref{quater_conj}. These fluxes have logarithmic interaction with each other which is mediated by the spin-wave magnons. Now consider quantum disordering this phase \textit{without proliferating these defects}. The resulting phase would be a paramagnet that would correspond to the deconfined phase of a pure quaternion gauge theory. The excitations in this phase would correspond to gapped spinons, quaternion fluxes and their composites (`dyons') which would have interesting fusion and braiding properties. Additionally, the ground state of this phase would have topological degeneracy on a torus.

The topological properties of this phase are well understood and a detailed discussion of the particles and their topological interactions was provided by Propitius and Bais in reference \cite{bais95}. The distinct magnetic fluxes are in one-to-one correspondence with the topological defects of the nematic phase, hence they are classified by the conjugacy classes of the quaternion group listed in eqn. \ref{quater_conj}. As in any gauge theory, the electric charges are labelled by the irreducible representations of the gauge group. In particular, the spinons $z_{\uparrow}, z_{\downarrow}$ transform under the two-dimensional representation of the quaternion group while the flux composites of the form $|i \sigma_a, -i\sigma_a>$ with zero net flux  also carry charge and transform under the one-dimensional representation of the quaternion group. Finally, one could have dyonic particles which correspond to the bound state of a flux and an electric charge. The dyons corresponding to a flux $\Phi$ are classified by the irreducible representations of the centralizer corresponding to conjugacy class of $\Phi$ (one may recall that the centralizer of a conjugacy class is the set of elements which commute with all elements of that conjugacy class). Since the centralizer of the $\bar{C_0}$ conjugacy class is the whole group $Q$, the bound state of a spinon and a $\bar{C_0}$ flux is classified by the irreducible representations of $Q$. On the other hand, the centralizer of the $C_x, C_y$ and $C_z$ conjugacy classes is the $Z_4$ group, hence the corresponding dyons fall into four distinct classes corresponding to the four one-dimensional representations of the $Z_4$ group.

Next consider the ground state degeneracy on a torus. Since the deconfined phase is gapped, one could calculate the degeneracy on a torus by restricting the system to a single plaquette with periodic boundary conditions. Thus, the number of ground states is proportional to the number of inequivalent flux configurations that satisfy $\prod_{\Box}\, \sigma_{ij} \sigma_{jk} \sigma_{kl} \sigma_{li} = 1$. A simple counting shows that this number is 22. As is well known, the ground state degeneracy for a topological phase is equal to the number of distinct particle excitations of the system. Thus there must be 22 particles in a quaternion gauge-matter theory as could be easily verified by direct calculation \cite{bais95}.

Since spinons transform under a two-dimensional representation of the quaternion group, it's also interesting to consider the effect of transportation of a spinon around a magnetic flux. For example, when a spinon $z$ carrying spin-up, $ z \equiv \left[ 1 \,\, \, 0 \right]^T$, goes around a flux $C_x$, it is transformed to $z' = \sigma_x \, z = \left[ 0 \,\, \, 1 \right]^T$ i.e. a spinon carrying down-spin. This may seem counterintuitive as it seemingly violates global spin-conservation. Since the ground state is a spin-singlet, one needs to be careful while discussing individual spinons. An instructive thought experiment which resolves this paradox is the following. Consider putting our system on a torus with a $\sigma_x$ vortex threading through one of the holes of the torus. Next an up spinon, down spinon pair is created out of the vacuum and the up-spinon is transported around the vortex keeping the location of the down spinon fixed. Finally, the up spinon is brought back to its original location. The interesting question is whether the spinon pair remain in the spin singlet sector (and hence can be annihilated) at the end of this process. 

To answer this, we note that the gauge invariant wave-function for the pair is given by $|\psi\rangle  = z^\dagger_{\bm{r} \uparrow} q_{\bm{r\, r_1}} q_{\bm{r_1 \, r_2}} ... q_{\bm{r_j \,r'}} z^\dagger_{\bm{r'} \downarrow}|0\rangle$ where $\bm{r, r'}$ are the locations of up and down spinon respectively and $q_{\bm{r_n r_{n+1}}}$'s are the values of the gauge fields on the links connecting the two spinons. For concreteness, let us assume that a $C_x$ vortex threads along the $x$-axis and is parameterized by the gauge choice $q_{\bm{r} \bm{r'}} = i \sigma_x $ for $\bm{r} = (0, r \,\hat{x}) \,, \bm{r'} = \bm{r} + \hat{y} \,\,\, \forall\, r $ while $q_{\bm{r} \bm{r'}} = 1$ otherwise. As the up spinon passes through the $x$-axis, the field operator $z_\uparrow$ transforms to $z_\downarrow$ while  the string operator  $\mathcal{S} = q_{\bm{r\, r_1}} q_{\bm{r_1 \, r_2}} ... q_{\bm{r_j \,r'}}$  connecting the two spinons transforms as $\mathcal{S} \rightarrow \sigma_x \mathcal{S} = \sigma_x $ since in the ground state the only contribution to $\mathcal{S}$ comes from the vortex gauge field. This implies that at the end of the adiabatic transformation, the up-spinon becomes a down-spinon and vice-versa since the operator $\mathcal{S}$ acts on the down-spinon and transforms it to an up-spinon. Hence the spinon pair behave like an EPR spin-singlet \cite{einstein} and the spin is delocalized while they are separated.

Having described various possible phases of the action $S$, it is worthwhile to compare our approach to standard methods for obtaining spin-liquid states. Our current understanding of spin-liquids with bosonic spinons draws heavily on the slave-boson formulation. In this approach, one represents the spin-$S$ operator in terms of a two component boson $b = \left[ b_1 \, \, b_2 \right]^{T}$ as $\bm{S} = \frac{1}{2}b^{\dagger} \bm{\sigma} b$. The Hilbert space of the bosons is projected to the physical subspace of a spin by the relation $b^{\dagger} b = 2 S$. Clearly, all physical operators are invariant under the $U(1)$ local transformation, $b(\bm{r}) \rightarrow e^{i\phi(\bm{r})} (\bm{r})$. The physical symmetries of the underlying spin-Hamiltonian are realized as transformation of spinons $b$ under  these symmetries  combined with a local $U(1)$ transformation. The spin-liquids correspond to states that do not break any physical symmetry and their low energy description consists of spinons $b$ coupled to a gauge field whose gauge group is always a subgroup of $U(1)$. 

Could the non-abelian quaternion phase described in this paper be described by such a projective construction? Interestingly, the answer is no! First and foremost, as we mentioned above the standard Schwinger boson technique can only describe abelian spin-liquids. Further, the spin-operator cannot be written as an operator quadratic in spinons $z$ for the following reason. The nematic order parameter $Q_{\mu \nu}$ is quartic in $z$'s and since it transforms as a spin-two operator, one needs to take it's tensor product with itself to construct a gauge invariant physical spin operator $\bm{S}$. Explicitly, $\bm{S}$ is given by $S_a = -i \epsilon^{abc} Q_{bd} Q_{dc}$. Therefore $\bm{S}$ would be of degree eight in $z$'s, which is very different than the usual slave boson theories. 

The phase transition between the three sublattice nematic phase and the non-abelian spin-liquid has also many interesting features. This phase transition corresponds to Higgs transition for the spinons $z$ and as we argue, is not describable within Landau-Ginzburg-Wilson paradigm. Let us start by analyzing the symmetries of the critical action. The symmetry under physical spin rotation corresponds to the left multiplication of the matrix $U$ (defined in the eqn. \ref{defn_U}) by an $SU(2)$ matrix. Remarkably, the action is invariant even under the \textit{right} multiplication of $U$ by an $SU(2)$ matrix and thus the critical theory has in fact $O(4) \sim SU(2) \times SU(2)$ symmetry! This is because the under unit translation $T_{\bm{\hat{a}}}$ on the triangular lattice, the $\bm{n_a}$ transform as:

\begin{equation}
 T_{\bm{\hat{a}}}: \,\, \bm{n_1} \rightarrow \bm{n_2}, \,\, \bm{n_2} \rightarrow \bm{n_3}, \,\, \bm{n_3} \rightarrow \bm{n_1}
\end{equation}

One finds that the operation of right multiplication of $U$ by an $SU(2)$ matrix corresponds to the rotation of $n_a$'s amongst each other. The generators of these rotations $K_a$ (a = 1, 2, 3) satisfy

\begin{equation}
 \left[\bm{n_a},K_b\right] = i\epsilon_{abc} \bm{n_c}
\end{equation}

Since the critical action must have both spin-rotation and translational invariance, it would be invariant under the transformation $U \rightarrow V_R U V_L$ where $V_R, V_L \in SU(2)$. Hence the critical theory is

\begin{eqnarray}
S & = & \frac{1}{g} \int d^2x \,\, Tr \left[ (\partial_\mu U^{\dagger}) \, (\partial_\mu U)\right] \\ \nonumber
& = & \frac{1}{g} \int d^2x \,d \tau \, |\partial_\mu z_\alpha|^2
\end{eqnarray}

with $z^{\dagger} z = 1$. We emphasize that the critical theory can be written in terms of spinons $z$ only because the topological defects are suppressed which renders them single-valued. We note that the critical theory is very similar to that for the phase transition between a spiral anti-ferromagnet and a $Z_2$ spin-liquid in a spin $S = 1/2$ triangular lattice magnet \cite{senthil94}.

The fact that that the nematic order-parameter $Q_{\mu \nu}$ is of fourth degree in $z$ has dramatic consequences for the critical correlations. For example, the critical exponent $\overline{\eta}$ defined by

\begin{equation}
 \left\langle Q_{\mu \nu} (k, \omega) Q_{\mu \nu} (k, \omega)\right\rangle  \sim \frac{1}{(\omega^2 - k^2)^{1-\overline\eta/2}} \label{scal_op}
\end{equation}

would have a large value which equals $\eta = 3$ in a large $N$ limit if one generalizes the $O(4)$ model to an $O(N)$ model. For finite $N$, one would obtain $\eta > 3$ whose precise numerical value we do not calculate here. This is very large compared to the anomalous exponents corresponding to the order parameter in usual Landau-Ginzburg-Wilson theories. 

Apart from the order parameter, the six conserved currents associated with the $O(4)$ symmetry would also have power-law correlations at the critical point. Three of these, the conserved total spin $\bm{S_{tot}} = \sum_{\bm{r}} {\bm{S(r)}}$ are conserved microscopically while the other three are the $K_a$'s defined above which are conserved only the low energy effective theory.

The conserved currents acquire no anomalous dimensions and hence have scaling dimension $d = 2$. Therefore their correlations at the critical point are given by:

\begin{equation}
 \left\langle J_a(\bm{r},\tau) \, J_a(\bm{0},0)\right\rangle \sim \frac{1}{(r^2 + \tau^2)^2} \label{scal_conserved}
\end{equation}

where $\bm{J} \equiv \left\lbrace \bm{S}, \, \bm{K}\right\rbrace$. Comparing eqn. \ref{scal_op} with eqn. \ref{scal_conserved}, one notices that the conserved currents have a \textit{slower} decay than that for the order parameter which is rather unusual. 

One may ask what interactions one might add in the Hamiltonian $H$ so as to destroy the nematic state? One simple way to obtain a paramagnet state out of the nematic state is to consider a stacked triangular lattice spin $S = 1$ system and add an antiferromagnet inter-layer interaction $J_{\perp}$. When $J_{\perp} \gg J_{\parallel}, K_{\parallel}$, the intra-layer couplings, one would obtain a translationally invariant paramagnet state corresponding to decoupled spin $S = 1$ chains. On the other hand, when $J_{\perp} \ll J_{\parallel}, K_{\parallel}$, one obtains ordered three-sublattice nematic state. Thus one can quantum disorder the nematic state by tuning $J_{\perp}/J_{\parallel}, J_{\perp}/K_{\parallel}$. It would be interesting to explore the possibility of realizing stacked copies of the non-abelian spin-liquid described in this paper in such a setting.

In summary, we have described a non-abelian spin-liquid in a spin $S = 1$ quantum magnet on triangular lattice which can not be accessed within the standard slave boson/fermion projective construction. The non-abelian phase has interesting topological features captured by the ground state degeneracy on a torus and braiding and fusion of its excitations. We also described a non-Landau phase transition between this spin-liquid and a non-collinear nematic state. The nematic correlations near this phase transition are characterized by large anomalous dimension.

\textit{Acknowledgements}: We would like to thank Michael Levin and Ashvin Vishwanath for helpful discussions. TS was supported by NSF Grant DMR-0705255.

\end{document}